\input harvmac
\input amssym

\def\unit{\relax{\rm 1\kern-.26em I}}
\def\nada{\relax{\rm 0\kern-.30em l}}
\def\tilde{\widetilde}



\noblackbox
\def\IL{\relax{\rm I\kern-.18em L}}
\def\IH{\relax{\rm I\kern-.18em H}}
\def\IR{\relax{\rm I\kern-.18em R}}
\def\IC{\relax\hbox{$\inbar\kern-.3em{\rm C}$}}
\def\IZ{\relax\ifmmode\mathchoice
{\hbox{\cmss Z\kern-.4em Z}}{\hbox{\cmss Z\kern-.4em Z}}
{\lower.9pt\hbox{\cmsss Z\kern-.4em Z}} {\lower1.2pt\hbox{\cmsss
Z\kern-.4em Z}}\else{\cmss Z\kern-.4em Z}\fi}

\def\CN {{\cal N}}
\def\CR {{\cal R}}

\def\CL {{\cal L}}

\def\CO {{\cal O}}

\def\CG {{\cal G}}

\def\CA{{\cal A}}


\def\CN {{\cal N}}

\def\CO {{\cal O}}

\font\manual=manfnt \def\dbend{\lower3.5pt\hbox{\manual\char127}}

\def\IZ{\relax\ifmmode\mathchoice
{\hbox{\cmss Z\kern-.4em Z}}{\hbox{\cmss Z\kern-.4em Z}}
{\lower.9pt\hbox{\cmsss Z\kern-.4em Z}} {\lower1.2pt\hbox{\cmsss
Z\kern-.4em Z}}\else{\cmss Z\kern-.4em Z}\fi}

\def\bar{\overline}

\def\rt2{\sqrt{2}}
\def\irt2{{1\over\sqrt{2}}}

\def\hat{\widehat}
\def\slashchar#1{\setbox0=\hbox{$#1$}           
   \dimen0=\wd0                                 
   \setbox1=\hbox{/} \dimen1=\wd1               
   \ifdim\dimen0>\dimen1                        
      \rlap{\hbox to \dimen0{\hfil/\hfil}}      
      #1                                        
   \else                                        
      \rlap{\hbox to \dimen1{\hfil$#1$\hfil}}   
      /                                         
   \fi}

\def\foursqr#1#2{{\vcenter{\vbox{
    \hrule height.#2pt
    \hbox{\vrule width.#2pt height#1pt \kern#1pt
    \vrule width.#2pt}
    \hrule height.#2pt
    \hrule height.#2pt
    \hbox{\vrule width.#2pt height#1pt \kern#1pt
    \vrule width.#2pt}
    \hrule height.#2pt
        \hrule height.#2pt
    \hbox{\vrule width.#2pt height#1pt \kern#1pt
    \vrule width.#2pt}
    \hrule height.#2pt
        \hrule height.#2pt
    \hbox{\vrule width.#2pt height#1pt \kern#1pt
    \vrule width.#2pt}
    \hrule height.#2pt}}}}
\def\psqr#1#2{{\vcenter{\vbox{\hrule height.#2pt
    \hbox{\vrule width.#2pt height#1pt \kern#1pt
    \vrule width.#2pt}
    \hrule height.#2pt \hrule height.#2pt
    \hbox{\vrule width.#2pt height#1pt \kern#1pt
    \vrule width.#2pt}
    \hrule height.#2pt}}}}
\def\sqr#1#2{{\vcenter{\vbox{\hrule height.#2pt
    \hbox{\vrule width.#2pt height#1pt \kern#1pt
    \vrule width.#2pt}
    \hrule height.#2pt}}}}

\def\figin{\epsfcheck\figin}\def\figins{\epsfcheck\figins}
\def\epsfcheck{\ifx\epsfbox\UnDeFiNeD
\message{(NO epsf.tex, FIGURES WILL BE IGNORED)}
\gdef\figin##1{\vskip2in}\gdef\figins##1{\hskip.5in}
\else\message{(FIGURES WILL BE INCLUDED)}%
\gdef\figin##1{##1}\gdef\figins##1{##1}\fi}
\def\DefWarn#1{}
\def\figinsert{\goodbreak\midinsert}
\def\ifig#1#2#3{\DefWarn#1\xdef#1{fig.~\the\figno}
\writedef{#1\leftbracket fig.\noexpand~\the\figno}%
\figinsert\figin{\centerline{#3}}\medskip\centerline{\vbox{\baselineskip12pt
\advance\hsize by -1truein\noindent\footnotefont{\bf
Fig.~\the\figno:\ } \it#2}}
\bigskip\endinsert\global\advance\figno by1}

\lref\CallanZE{
  C.~G.~.~Callan, S.~R.~Coleman and R.~Jackiw,
  ``A New improved energy - momentum tensor,''
  Annals Phys.\  {\bf 59}, 42 (1970).
}

\lref\FerraraPZ{
  S.~Ferrara and B.~Zumino,
  ``Transformation Properties Of The Supercurrent,''
  Nucl.\ Phys.\  B {\bf 87}, 207 (1975).
}

\lref\GrinsteinQK{
  B.~Grinstein, K.~A.~Intriligator and I.~Z.~Rothstein,
  ``Comments on Unparticles,''
  Phys.\ Lett.\  B {\bf 662}, 367 (2008)
  [arXiv:0801.1140 [hep-ph]].
}

\lref\PolchinskiDY{
  J.~Polchinski,
  ``Scale and Conformal Invariance in Quantum Field Theory,''
  Nucl.\ Phys.\  B {\bf 303}, 226 (1988).
}

\lref\ZamolodchikovGT{
  A.~B.~Zamolodchikov,
  ``Irreversibility of the Flux of the Renormalization Group in a 2D Field
  Theory,''
  JETP Lett.\  {\bf 43}, 730 (1986)
  [Pisma Zh.\ Eksp.\ Teor.\ Fiz.\  {\bf 43}, 565 (1986)].
}

\lref\KomargodskiRB{
  Z.~Komargodski and N.~Seiberg,
  ``Comments on Supercurrent Multiplets, Supersymmetric Field Theories and
  Supergravity,''
  JHEP {\bf 1007}, 017 (2010)
  [arXiv:1002.2228 [hep-th]].
}

\lref\SeibergAJ{
  N.~Seiberg and E.~Witten,
  ``Monopoles, duality and chiral symmetry breaking in N=2 supersymmetric
  QCD,''
  Nucl.\ Phys.\  B {\bf 431}, 484 (1994)
  [arXiv:hep-th/9408099].
}

\lref\ArgyresXN{
  P.~C.~Argyres, M.~Ronen Plesser, N.~Seiberg and E.~Witten,
  ``New N=2 Superconformal Field Theories in Four Dimensions,''
  Nucl.\ Phys.\  B {\bf 461}, 71 (1996)
  [arXiv:hep-th/9511154].
}

\lref\GaiottoWE{
  D.~Gaiotto,
  ``N=2 dualities,''
  arXiv:0904.2715 [hep-th].
}

\lref\SeibergPQ{
  N.~Seiberg,
  ``Electric - magnetic duality in supersymmetric nonAbelian gauge theories,''
  Nucl.\ Phys.\  B {\bf 435}, 129 (1995)
  [arXiv:hep-th/9411149].
}

\lref\DorigoniRA{
  D.~Dorigoni and V.~S.~Rychkov,
  ``Scale Invariance + Unitarity $=>$ Conformal Invariance?,''
  arXiv:0910.1087 [hep-th].
}

\lref\KachruYH{
  S.~Kachru, X.~Liu and M.~Mulligan,
  ``Gravity Duals of Lifshitz-like Fixed Points,''
  Phys.\ Rev.\  D {\bf 78}, 106005 (2008)
  [arXiv:0808.1725 [hep-th]].
}

\lref\HoravaUW{
  P.~Horava,
  ``Quantum Gravity at a Lifshitz Point,''
  Phys.\ Rev.\  D {\bf 79}, 084008 (2009)
  [arXiv:0901.3775 [hep-th]].
}

\lref\NakayamaWX{
  Y.~Nakayama,
  ``Higher derivative corrections in holographic Zamolodchikov-Polchinski
  theorem,''
  arXiv:1009.0491 [hep-th].
}

\lref\MackJE{
  G.~Mack,
  ``All Unitary Ray Representations Of The Conformal Group SU(2,2) With
  Positive Energy,''
  Commun.\ Math.\ Phys.\  {\bf 55}, 1 (1977).
}

\lref\CardyCWA{
  J.~L.~Cardy,
  ``Is There a c Theorem in Four-Dimensions?,''
  Phys.\ Lett.\  B {\bf 215}, 749 (1988).
}

\lref\ElShowkGZ{
  S.~El-Showk, Y.~Nakayama and S.~Rychkov,
  ``What Maxwell Theory in $D<>4$ teaches us about scale and conformal
  invariance,''
  arXiv:1101.5385 [hep-th].
}

\lref\KomargodskiPC{
  Z.~Komargodski and N.~Seiberg,
  ``Comments on the Fayet-Iliopoulos Term in Field Theory and Supergravity,''
  JHEP {\bf 0906}, 007 (2009)
  [arXiv:0904.1159 [hep-th]].
}

\lref\JackiwVZ{
  R.~Jackiw, S.~-Y.~Pi,
  ``Tutorial on Scale and Conformal Symmetries in Diverse Dimensions,''
J.\ Phys.\ A {\bf A44}, 223001 (2011).
[arXiv:1101.4886 [math-ph]].
}

\lref\deWitFH{
  B.~de Wit, M.~Rocek,
  ``Improved Tensor Multiplets,''
Phys.\ Lett.\  {\bf B109}, 439 (1982).
}

\lref\PonsNB{
  J.~M.~Pons,
  ``Noether symmetries, energy-momentum tensors and conformal invariance in classical field theory,''
[arXiv:0902.4871 [hep-th]].
}

\lref\MagroAJ{
  M.~Magro, I.~Sachs, S.~Wolf,
  ``Superfield Noether procedure,''
Annals Phys.\  {\bf 298}, 123-166 (2002).
[hep-th/0110131].
}

\lref\SiegelAI{
  W.~Siegel,
  ``Gauge Spinor Superfield as a Scalar Multiplet,''
Phys.\ Lett.\  {\bf B85}, 333 (1979).
}

\rightline{CERN-PH-TH/2011-002}
\Title{\vbox{\baselineskip12pt }} {\vbox{\centerline{On R-symmetric Fixed Points and Superconformality}}}
\smallskip
\centerline{Ignatios Antoniadis\foot{On leave from CPHT (UMR CNRS 7644) Ecole Polytechnique, F-91128 Palaiseau.} and Matthew Buican}
\smallskip
\bigskip
\centerline{{\it Department of Physics, CERN Theory Division, CH-1211 Geneva 23, Switzerland}} %
\vskip 1cm

\noindent An important unanswered question in quantum field theory is to understand precisely under which conditions scale invariance implies invariance under the full conformal group. While the general answer in two dimensions has been known for over 20 years, a precise nonperturbative relation between scale and conformal invariance in higher dimensions has been lacking. In this note, we specialize to four dimensions and give a full quantum mechanical proof
that certain unitary R-symmetric fixed points are necessarily superconformal. Among other consequences, this result implies that the infrared fixed points of $\CN=1$ supersymmetric quantum chromodynamics are superconformal.

\bigskip
\Date{February 2011}

\newsec{Introduction}
Often, when discussing systems at a fixed point of the renormalization group (RG), we simply assume that the theory is invariant under the full conformal group. This assumption is backed up by the apparent argument that scale invariance implies vanishing of the theory's beta functions, which in turn implies vanishing of the trace of the stress tensor, which then implies conformal invariance.

This chain of logic has at least two gaps. First, there is an ambiguity in the definition of the stress tensor. Indeed, a theory has a whole family of stress tensors related by improvement transformations, and each member of the family has, in general, a different trace. However, the real gap in the above argument is that scale invariance does not, a priori, imply vanishing of the beta functions of the theory.

It is easy to see why the relation between scale invariance and the value of the beta functions is actually more subtle. Indeed, one can imagine following a theory along its Wilsonian RG trajectory by integrating out lower and lower momentum degrees of freedom. As we follow the theory along its flow, the various couplings of the theory change by different amounts depending on their respective beta functions, and it may happen that these changes eventually conspire to give a term that is actually a total derivative. In such a case, we would arrive at a fixed point---a point of exact scale invariance---with nonvanishing beta function(s). It may then happen that the trace of the stress tensor cannot be improved away, and so full conformal invariance is not respected.

In two dimensions, Zamolodchikov \ZamolodchikovGT\ and Polchinski \PolchinskiDY\ showed that such fixed points cannot exist. More precisely, they proved that a scale and Poincar\'e invariant unitary quantum field theory (QFT) with a discrete operator spectrum
is necessarily a conformal field theory. As an aside, we note that while the original proof of the Zamolodchikov-Polchinski (ZP) theorem follows from the existence of a \lq\lq$c$-function" that decreases along the RG flow and interpolates between the central charges of the ultraviolet (UV) and infrared (IR) fixed points, the two results seem---at least superficially---to be logically independent \DorigoniRA.

In higher dimensions, the situation is far less clear. While there are no known unitary examples of higher dimensional nonconformal fixed points, there is also no general proof that they do not exist.\foot{Often in the QFT literature, scale invariance, Poincar\'e invariance, and unitarity are claimed to imply conformality even in higher dimensions. Typically, the result of \CallanZE\ is cited as proving this claim in four dimensions. However, the authors of \CallanZE\ only proved this statement at the classical level.} \foot{Relaxing the assumption of Poincar\'e invariance, one finds the famous Lifshitz-like fixed points, which are generally only scale invariant. Such fixed points have received much attention recently in the quantum gravity literature \KachruYH\ \HoravaUW.} \foot{The authors of two interesting recent papers \JackiwVZ\ElShowkGZ\ found that free Maxwell theories in $d\ne4$ are scale invariant but non-conformal. However, these examples do not contain well-defined scaling currents. See also the discussion in \PonsNB. It may be that even in these other (integer) dimensions, theories with well-defined scaling currents are necessarily conformal.} In his seminal paper \PolchinskiDY, Polchinski argued that certain classes of four dimensional fixed points are necessarily conformal. In particular, he showed that the perturbative Banks-Zaks (BZ) fixed points of quantum chromodynamics (QCD) and the Wilson-Fisher (WF) fixed point of $\phi^4$ theory in $4-\epsilon$ dimensions are conformal. However, nothing is known in general about the fixed point behavior of more complicated theories. Indeed, even for simple multiflavor generalizations of $\phi^4$ theory, it is only known that scale invariance implies conformal invariance at one loop \PolchinskiDY. Recently, Dorigoni and Rychkov \DorigoniRA\ have extended these results and shown that adding fermionic flavors still does not allow for nonconformal fixed points at one loop.

Clearly, the challenge is to develop a nonperturbative understanding of scale versus conformal invariance in higher dimensions.\foot{Recently, Nakayama \NakayamaWX\ has used gauge / gravity duality to suggest that scale invariance implies conformal invariance for the class of theories that have a gravitational dual.} The reason for this is that unlike the anomalies of global internal symmetries, the anomalies of the scale and special conformal transformations are not saturated at finite loop order. Therefore, one needs a nonperturbative understanding even for perturbative fixed points (never mind for strongly-coupled ones).

Although interacting scale-invariant theories in four dimensions were once thought to be rather rare, a remarkable number of nontrivial fixed-point theories have been discovered in recent times. Most of these examples are supersymmetric (SUSY) and include, among others, $\CN=4$ Super Yang-Mills, certain $\CN=2$ super QCD (SQCD) theories \SeibergAJ\ArgyresXN\ and generalizations thereof \GaiottoWE, as well as the IR fixed points of particular $\CN=1$ SQCD theories \SeibergPQ. While some of these theories are known to be invariant under the full superconformal group, others are not.

In this paper we specialize to four-dimensional R-symmetric theories with $\CN=1$ supersymmetry (SUSY) and show that any unitary fixed point in this class of theories is either a superconformal field theory (SCFT) or has at least two real nonconserved dimension two scalar singlet operators (SSOs).\foot{These are Lorentz scalar operators, $S_i$, that are singlets under all the symmetries of the theory,
 satisfy $\left\{Q_{\alpha}, \left[Q_{\beta}, S_i\right]\right\}\ne0$, and are not related by the addition of a conserved real singlet operator, $J$ (i.e., an operator satisfying $\left\{Q_{\alpha}, \left[Q_{\beta}, J\right]\right\}=0$).} \foot{In deriving this statement, we study theories with well-defined (gauge-invariant and local) dilatation currents. Relaxing this requirement, we note that a restricted version of the free two-form theory is $R$-symmetric and scale-invariant but non-conformal, thus generalizing the discussion in \PonsNB\ to the SUSY case. Similar comments apply to scalars with shift symmetry.} We then generalize this result and show that any fixed point arising as the IR limit of a UV SCFT is necessarily superconformal provided that three conditions hold. First, we assume that the deformation that starts the flow is  a marginally relevant deformation that leads to only one nonconserved SSO (modulo deformations by conserved SSO operators) of dimension two in the UV.\foot{When we say that a nonconserved SSO has dimension two in the UV, we mean that it has dimension two before turning on the marginally relevant deformation (i.e., it descends from a dimension two SSO operator in the UV SCFT). As we will see below, using the $R$ symmetry current superfield, we can make this notion precise in the cases of interest to us.} Second, we assume that the IR R-symmetry corresponds to a nonanomalous symmetry of the flow. Finally, we assume that the theory has a well-defined Ferrara-Zumino (FZ) multiplet. While such fixed points may not be the most general ones, many of the fixed points mentioned in the previous paragraph are either known to be conformal or fall into this broad category. As we will see below, even though the theories we consider can be extremely intricate, SUSY and R-symmetry give us a strong handle on their possible fixed point behavior. Indeed, we will see that by imposing unitarity constraints on the dilatation current multiplet of these theories, we will be able to rule out even strongly-coupled nonsuperconformal fixed points in the deep IR under the conditions described above.

The plan of this paper is as follows. In the next section we review some basic aspects of the conformal group as well as a necessary and sufficient condition for the existence of a nonconformal (not necessarily SUSY) fixed point due to Polchinski. We then review unitarity bounds on operators in conformal and nonconformal fixed-point theories. In the third section we specialize to R-symmetric SUSY theories and review the multiplet for the corresponding conserved R-symmetry current. We then use this superfield to construct the multiplet for the conserved dilatation current, and we impose unitarity on this multiplet. In the following subsection, we use the dilatation current multiplet to show that R-symmetric fixed points are either superconformal or have more than one independent real nonconserved dimension two SSO. Then, by constraining the UV behavior of the dilatation current superfield under marginally relevant deformations of the UV fixed-point theory that lead to only one dimension two SSO (modulo deformations by conserved SSOs), we show that the corresponding class of IR theories is necessarily superconformal. Along the way, we apply this discussion to the conformal window of $\CN=1$ SQCD. Throughout the paper, we assume that the scaling of the operators in the supercurrent multiplet is canonical, and in an appendix we justify this claim. We conclude with a brief discussion of open questions.

\newsec{Scale versus conformal invariance}
In four dimensions, the conformal group consists of 15 generators satisfying an algebra isomorphic to $SO(4,2)$. These generators consist of the four translations, $P_{\mu}$, the six Lorentz rotations, $\omega_{\mu\nu}$, the dilatation generator, $\Delta$, and the four special conformal transformations, $K_{\mu}$. Since the generators of the conformal group transform with definite scaling dimensions, it follows that for any subgroup, $\CG\subset SO(4,2)$, the set $\CG\cup\left\{\Delta\right\}$ is also a subgroup of the conformal group. In particular, the Poincar\'e group plus the dilatations is a subgroup.

As a result, at the level of the charge algebra, it is perfectly consistent to imagine a Poincar\'e invariant QFT that is scale-invariant but not invariant under the special conformal transformations.

\subsec{Polchinski's criterion for a nonconformal fixed point}
Any theory that is invariant under scalings must have a corresponding dimension three conserved current, $\Delta_{\mu}$. On general grounds, $\Delta_{\mu}$ must have the following form
\eqn\Dmugen{
\Delta_{\mu}=x^{\nu}T_{\mu\nu}+\CO_{\mu},
}
where $T_{\mu\nu}$ is the stress tensor of the theory and $\CO_{\mu}$ is an internal current (sometimes referred to as the \lq\lq virial" current) that does not explicitly depend on the space-time coordinates. Conservation of $\Delta_{\mu}$ implies that
\eqn\Dmucond{
T\equiv T^{\mu}_{\ \mu}=-\partial^{\mu}\CO_{\mu}.
}
Clearly if $\CO_{\mu}$ is conserved, then $T=0$, and the theory is conformal.\foot{Recall that a theory is conformal if and only if it is possible to improve the stress tensor such that it is traceless.}

More generally, Polchinski \PolchinskiDY\ showed that one can improve $T_{\mu\nu}$ and define a new stress tensor $T'_{\mu\nu}$ such that
\eqn\impT{
T'^{\mu}_{\ \ \mu}=0,
}
if and only if there exists a local gauge-invariant $L_{\mu\nu}$ such that
\eqn\COconf{
\CO_{\mu}=j_{\mu}+\partial^{\nu}L_{\mu\nu}, \ \ \ \partial^{\mu}j_{\mu}=0.
}
For a theory that is classically scale-invariant, it is always possible to construct such an improved traceless tensor $T'_{\mu\nu}$ \CallanZE, and so it follows that such a theory is necessarily conformal.

At the quantum level, this claim has never been proven in general, but any nonconformal fixed-point QFT must have a local gauge-invariant dimension three operator not of the form \COconf\ that satisfies \Dmucond. It is then straightforward to see that the Wilson-Fisher fixed point for $\phi^4$ theory is conformal since we must have
\eqn\COmuphifour{
\CO_{\mu}=c\ \partial_{\mu}\left(\phi^2\right),
}
where $c$ is a constant. Similarly, for the BZ fixed points of QCD, there is no BRST nontrivial candidate for $\CO_{\mu}$, and so such fixed points are also conformal.

While these results are interesting, they are rather difficult to generalize. For example, if we consider quartic theories with $N>1$ scalars, $\phi^i$, the corresponding broken $O(N)$ currents do not take the form \COconf, and so they are natural candidates to mix with $\CO_{\mu}$. Parameterizing this mixing in the most general way, we have
\eqn\ONcurr{
\CO_{\mu}=\kappa_{a}(\lambda)\ j_{\mu}^a,
}
where $\lambda$ is shorthand for the quartic couplings of the theory (we have suppressed the corresponding flavor indices), $a=1, ..., N(N-1)/2$ is an adjoint index for $O(N)$, and $j^a_{\mu}$ is the corresponding current. However, by imposing \Dmucond, Polchinski showed that $\kappa_{a}(\lambda)=0$ for the broken symmetries at one loop. Therefore, the broken $\CO(N)$ currents do not mix with the dilatation current at that order, and the fixed points are conformal at one loop. It is not known what happens at higher orders in perturbation theory.\foot{Using the same basic technique, Dorigoni and Rychkov \DorigoniRA\ added fermionic flavors, $\psi^i$, Yukawa interactions with the scalars, and showed that the one-loop fixed points are still conformal.}

\subsec{Unitarity bounds}
As we have seen above, the virial current, $\CO_{\mu}$, must have dimension three. In a general conformal field theory, there are well-known lower bounds on the quantum dimensions of local gauge-invariant operators \MackJE. Indeed, for a primary operator
\eqn\prim{
\CO_{(j_1, j_2)}, \ \ \ j_{1,2}\in SU(2)_{L, R},
}
unitarity implies that its dimension is bounded as follows
\eqn\CFTdimbd{\eqalign{
d\left(\CO_{(j_1,j_2)}\right)&\ge j_1+j_2+2, \ \ \ j_1\cdot j_2\ne0,\cr d\left(\CO_{(j_1,j_2)}\right)&\ge j_1+j_2+1, \ \ \ j_1\cdot j_2=0.
}}
Furthermore, an operator saturates these unitarity bounds if and only if it satisfies a simple differential equation; this requirement corresponds to the fact that a descendant operator is set to zero.

In the case of a nonconformal fixed point, the scaling dimensions of operators are still bounded from below \GrinsteinQK. These unitarity bounds are less familiar, but they will be crucial in what follows when we apply them to superpartners of the virial current $\CO_{\mu}$
\eqn\Scaleunit{
d\left(\CO_{(j_1,j_2)}\right)\ge j_1+j_2+1, \ \ \ \forall \ j_{1,2}.
}
Just as in the conformal case, a scalar operator, $\CO$, saturates \Scaleunit\ if and only if it satisfies a Klein-Gordon equation
\eqn\Scalarcond{
\partial^2\CO=0 \ \ \ \Leftrightarrow \ \ \ d\left(\CO\right)=1.
}
It also turns out that a spin $1/2$ operator, $\CO_{\dot\alpha}$, saturates \Scaleunit\ if and only if it satisfies a Dirac equation
\eqn\Diraccond{
\sigma^{\mu}_{\alpha\dot\alpha}\partial_{\mu}\CO^{\dot\alpha}=0 \ \ \ \Leftrightarrow \ \ \ d\left(\CO^{\dot\alpha}\right)=3/2.
}
Similarly, an antisymmetric tensor, $\CO_{[\mu\nu]}$, saturates \Scaleunit\ if and only if it satisfies a Maxwell equation
\eqn\Maxeqn{
\partial^{\nu}\CO_{[\nu\mu]}=0 \ \ \ \Leftrightarrow \ \ \ d\left(\CO_{[\nu\mu]}\right)=2.
}

In addition to these results, we will also need to consider spin $(3/2, 0)$ and $(1/2, 1)$ operators $\CO_{\alpha\beta\delta}$ and $\CO_{\dot\alpha\dot\beta\delta}$ saturating \Scaleunit\ with dimension $5/2$. As we will show in the appendix, unitarity forces such operators to satisfy
\eqn\fermunit{
\partial^{\nu}\sigma_{\nu\mu}^{\alpha\beta}\CO_{\alpha\beta\delta}=\partial^{\nu}\bar\sigma_{\nu\mu}^{\dot\alpha\dot\beta}\CO_{\dot\alpha\dot\beta\delta}=0.
}

\newsec{Superscale versus superconformal invariance}
In the rest of this paper, we will focus on SUSY fixed points. These theories must be invariant under a subgroup of the superconformal group that includes the dilatation generator and the super-Poincar\'e group. The superconformal group is much larger than the conformal group. In addition to the conformal group, the superconformal group contains a Weyl fermion of four supercharges, $Q_{\alpha}$. Closure then requires the existence of four additional dimension $-1/2$ supercharges, $S_{\alpha}$, for a total of eight fermionic generators. In the bosonic sector, closure requires, in addition to the 15 conformal generators discussed above an additional superconformal R-symmetry generator for a total of 16 bosonic charges. Note that since all the generators of the superconformal group transform with a definite dimension and R-charge, it follows that for any subgroup, $\CG$, of the superconformal group, $\CG\cup\left\{\Delta\right\}$ and $\CG\cup\left\{R\right\}$ are also subgroups.

Therefore, at the level of the charge algebra, one could imagine that a scale-invariant SUSY QFT is invariant under just the super-Poincar\'e group along with the dilatations and, possibly, an R-symmetry. In the remainder of this paper, we will study unitary fixed-point theories that possess an R-symmetry and show that under certain conditions, they are necessarily SCFTs.

\subsec{The R-current and dilatation current multiplets}
Since the theories we are studying have an R-symmetry, they have a corresponding dimension three conserved R-current. This current sits as the lowest component in a multiplet that contains the supercurrent and the stress tensor as well. Following the notation of \KomargodskiRB, we denote this multiplet $\CR_{\mu}$, and take it to satisfy
\eqn\defnR{\eqalign{
&\bar D^{\dot\alpha}\CR_{\alpha\dot\alpha}=\chi_{\alpha},\cr& \bar D_{\dot\alpha}\chi_{\alpha}=\bar D_{\dot\alpha}\bar\chi^{\dot\alpha}-D^{\alpha}\chi_{\alpha}=0,
}}
where $\CR_{\dot\alpha\alpha}\equiv-2\sigma^{\mu}_{\alpha\dot\alpha}\CR_{\mu}$. If the theory has an additional global symmetry, the corresponding current, $j_{\mu}$, transforms as the vector component of a real superfield, $J$, satisfying $D^2J=0$. This global symmetry can be combined with the R-symmetry in \defnR\ and leads to an ambiguity in $\CR_{\mu}$ described by \KomargodskiRB
\eqn\Ramb{\eqalign{
&\CR'_{\alpha\dot\alpha}=\CR_{\alpha\dot\alpha}+\left[D_{\alpha}, \bar D_{\dot\alpha}\right]J\cr&\chi'_{\alpha}=\chi_{\alpha}+{3\over2}\bar D^2D_{\alpha}J\cr& D^2J=0.
}}
As we will see momentarily, this shift affects the stress tensor and supercurrent through improvement transformations. In the case that 
\eqn\confcond{
\chi_{\alpha}=-{3\over2}\bar D^2D_{\alpha}J,
}
the theory can be improved so that $\chi_{\alpha}'=0$, and the theory is in fact superconformal. It then follows that the conserved R-current sitting at the bottom of $\CR'_{\mu}$ is just the Ferrara-Zumino R-current \FerraraPZ.

Solving \defnR\ in components, one finds
\eqn\Rsoln{\eqalign{
\CR_{\mu}&=j_{\mu}^R+i\theta^{\alpha}S_{\mu\alpha}-i\bar\theta_{\dot\alpha}\bar S_{\mu}^{\dot\alpha}+\theta\sigma^{\nu}\bar\theta\left(2T_{\mu\nu}+{1\over2}\epsilon_{\mu\nu\rho\sigma}\left(\partial^{\rho}j^{R\sigma}+{1\over4}F^{\rho\sigma}\right)\right)-{1\over2}\bar\theta^2\theta^{\beta}\sigma^{\nu}_{\beta\dot\beta}\partial_{\nu}\bar S_{\mu}^{\dot\beta}+\cr&{1\over2}\theta^2\bar\theta_{\dot\beta}\bar\sigma^{\nu\dot\beta\beta}\partial_{\nu}S_{\mu\beta}-{1\over4}\theta^4\partial^2j^R_{\mu}.
}}
The anomaly multiplet then takes the following form
\eqn\Chisoln{
\chi_{\alpha}=2i\sigma^{\mu}_{\alpha\dot\alpha}\bar S_{\mu}^{\dot\alpha}-\left(4T\delta^{\beta}_{\alpha}+iF_{\alpha}^{\ \beta}\right)\theta_{\beta}+2\theta^2\sigma^{\nu}_{\alpha\dot\alpha}\bar\sigma^{\mu\dot\alpha\beta}\partial_{\nu}S_{\mu\beta}+\CO(\theta\bar\theta),
}
where we have taken the $D$ component of $\chi_{\alpha}$ to satisfy
\eqn\Dcond{
D=-4T.
}

Let us note that for the theories that we are interested in there should be a well-defined dimension two real superfield $U$ such that
\eqn\FZwelldef{
\chi_{\alpha}=-{1\over4}\bar D^2D_{\alpha}U.
}
The fact that there is such a $U$ superfield corresponds to the fact that the FZ multiplet should be well-defined for scale-invariant theories. Indeed, the only type of theories known to have ill-defined FZ multiplets are theories with FI terms \KomargodskiPC\ and theories with nontrivial target space topology \KomargodskiRB. However, these theories manifestly break scale invariance.

Note that the ambiguity in \Ramb\ corresponds to the following improvement transformations of the component supercurrent and stress tensor\foot{In \Rsoln\ and \Chisoln, we choose our normalization conventions for $F_{\mu\nu}$ differently from those used in \KomargodskiRB.}
\eqn\improv{\eqalign{
&S_{\mu\alpha}'= S_{\mu\alpha}+2\left(\sigma_{\mu\nu}\right)_{\alpha}^{\ \beta}\partial^{\nu}J|_{\theta^{\beta}},\cr& T_{\mu\nu}'=T_{\mu\nu}+{1\over2}\left(\partial_{\mu}\partial_{\nu}-\eta_{\mu\nu}\partial^2\right)J|,
}}
where \lq $|_{\theta}$' and \lq $|$' denote the $\CO(\theta)$ and lowest order components of the $J$ superfield.

Let us now describe the multiplet for the dilatation current, $\Delta_{\mu}$. From \Dmugen\ and \Rsoln, we see that we can construct this multiplet as follows
\eqn\defnDelta{
\Delta_{\mu}=-{1\over8}x^{\nu}\bar\sigma_{\mu}^{\dot\alpha\alpha}\left[D_{\alpha}, \bar D_{\dot\alpha}\right]\CR_{\nu}+\CO_{\mu}=x^{\nu}\left(T_{\nu\mu}+{1\over4}\epsilon_{\nu\mu\rho\sigma}\left(\partial^{\rho}j^{R\sigma}+{1\over4}F^{\rho\sigma}\right)\right)+\CO_{\mu}|+...
}
The second term in this definition, $\CO_{\mu}$, is the virial current superfield and does not explicitly depend on the space-time coordinates.
\foot{Let us emphasize again that throughout this paper we study theories with well-defined (i.e., local and gauge-invariant) dilatation currents (thus allowing us to use our unitarity techniques in the following subsections). If we are willing to allow for a non-gauge-invariant dilatation current, then a restricted version of the free supersymmetric two-form theory is scale-invariant and $R$-symmetric but non-conformal (see \PonsNB\ for a non-SUSY discussion of scale versus conformal invariance for free $p$-form fields in various dimensions). Indeed, this theory is described by $\CL=\int d^4\theta G^2$, where $G=D^{\alpha}\psi_{\alpha}+\bar D_{\dot\alpha}\bar\psi^{\dot\alpha}$ contains in its $\theta\bar\theta$ component the (dual) field strength of the free two-form $b_{\mu\nu}$, i.e. $\epsilon_{\mu\nu\rho\sigma}h^{\nu\rho\sigma}$ \SiegelAI. Note that $\psi_{\alpha}$ is the supersymmetrization of the two-form and that the supergauge transformations take the form $\delta\psi_{\alpha}=i\bar D^2D_{\alpha}V$. These transformations leave $G$ invariant, and $G$ satisfies $D^2G=0$ by virtue of the fact that $\psi_{\alpha}$ is chiral.

In this theory, $R_{\alpha\dot\alpha}=c\cdot \ D_{\alpha}G\bar D_{\dot\alpha}G$, for $c\ne0$ an order one constant, and therefore $U=-c \cdot G^2$. Since $D^2U\ne0$, the theory is not conformal (see also \deWitFH\MagroAJ). However, the dilatation current is not gauge-invariant. To see this, one can use \Dcond\ and find that $T\supset c'\cdot \left(h_{\mu\nu\rho}\right)^2$ (for $c'\ne0$ an order one constant) is the only term in $T$ depending on the two-form. Therefore, the virial current depends on the two-form via $\CO_{\mu}\supset -3c'\cdot b^{\nu\rho}h_{\mu\nu\rho}$, and so the dilatation current is not invariant under the gauge transformations $\delta b_{\mu\nu}=\partial_{[\mu}\Lambda_{\nu]}$. However, if we restrict the gauge transformations to fall off to zero fast enough, then the dilatation charge is well-defined, and the theory is scale-invariant. Finally, let us again note that similar comments apply to a scalar with shift symmetry.

We will not treat theories of this type in the next section, and we will instead focus on theories with well-defined dilatation currents (a required condition for applying our unitarity arguments).} 

In order for $\Delta_{\mu}$ to admit an interpretation as the dilatation current mulitplet, it must be conserved. This fact implies the following (anomalous) conservation equation for the virial current
\eqn\consDelta{
\partial^{\mu}\CO_{\mu}=-{1\over8}D^{\alpha}\chi_{\alpha}.
}
Furthermore, it must be the case that the supercharge transforms with dimension $1/2$, and so\foot{Note that the supercharge does not act on the explicit factor of $x^{\nu}$ in \defnDelta\ since $x^{\nu}$ is a coordinate.}
\eqn\Deltarulei{
\left[Q_{\alpha}, \Delta\right]=\int d^3x\left[Q_{\alpha}, \Delta_0\right]=-{i\over2}Q_{\alpha}+\int d^3x\left(\CO_{0\alpha}-i(\sigma_0^{\ \mu})_{\alpha}^{\ \beta}S_{\mu\beta}+{i\over2}S_{0\alpha}\right)=-{i\over2}Q_{\alpha}.
}
In writing this equation, we have defined $\CO_{\mu\alpha}\equiv\left[Q_{\alpha}, \CO_{\mu}\right]$.\foot{In superfield notation, we have \eqn\COSF{\CO_{\mu}=\CO_{\mu}+i\theta^{\alpha}\CO_{\mu\alpha}-i\bar\theta_{\dot\alpha}\bar\CO_{\mu}^{\dot\alpha}+\theta\sigma^{\nu}\bar\theta\CO_{\mu\nu}+\CO(\theta^2, \bar\theta^2).}} From \Deltarulei, it follows that
\eqn\Ocond{
\CO_{\mu\alpha}={i\over2}\sigma_{\mu\alpha\dot\alpha}\bar\sigma^{\nu\dot\alpha\beta}S_{\nu\beta}+(\sigma_{\mu}^{\ \nu})^{\beta\delta}\partial_{\nu}\gamma_{\beta\delta\alpha}+(\bar\sigma_{\mu}^{\ \nu})^{\dot\beta\dot\delta}\partial_{\nu}\gamma_{\dot\beta\dot\delta\alpha}
}
where $\gamma_{\beta\delta\alpha}$ and $\gamma_{\dot\beta\dot\delta\alpha}$ are local, gauge-invariant dimension $5/2$ fermionic operators.

Now, from \consDelta\ and \Ocond, it is straightforward to see that the ambiguity in the R-symmetry \Ramb\ \improv\ corresponds to the following ambiguity in the virial current superfield\foot{There is an additional ambiguity in the definition of $\CO_{\mu}$ which corresponds to the ability to shift $\CO_{\mu}$ by a global conserved current.}
\eqn\viramb{
\CO'_{\mu}=\CO_{\mu}+{3\over2}\partial_{\mu}J.
}
Under such a transformation, $(\sigma_{\mu}^{\ \nu})^{\beta\delta}\partial_{\nu}\gamma_{\beta\delta\alpha}$ shifts as follows
\eqn\deformation{
(\sigma_{\mu}^{\ \nu})^{\beta\delta}\partial_{\nu}\gamma_{\beta\delta\alpha}'=(\sigma_{\mu}^{\ \nu})^{\beta\delta}\partial_{\nu}\gamma_{\beta\delta\alpha}+3\left(\sigma_{\mu}^{\ \nu}\right)_{\alpha}^{\ \beta}\partial_{\nu}j_{\beta}
}
Finally, let us note that it is clear that a transformation of the form \improv\ can be used to set $T=-\partial^{\mu}\CO_{\mu}=0$ if and only if
\eqn\SCCOcond{
\CO_{\mu}=\tilde\CO_{\mu}+\partial_{\mu}\hat J,
}
for real $\tilde\CO_{\mu}$ and $\hat J$ satisfying $\partial^{\mu}\tilde\CO_{\mu}=D^2\hat J=0$.

\subsec{The consequences of unitarity on the current multiplets}
In the previous section, we imposed two consistency conditions on the dilatation multiplet, $\Delta_{\mu}$. The first condition was current conservation, which led to the superfield Eq. \consDelta\ for the virial current superfield. The second condition was that the supercharge transforms with dimension $1/2$. This constraint led us to the condition \Ocond\ on the dimension $7/2$ fermionic partner of the virial current.

In this section, we will impose unitarity and closure of the SUSY algebra on $\CO_{\mu}$. First, consider imposing unitarity on \Ocond. Using the unitarity condition in \fermunit, we see that the contributions of $\gamma_{\dot\beta\dot\delta\alpha}$ and the higher spin components of $\gamma_{\beta\delta\alpha}$ vanish, leaving
\eqn\Ocond{
\CO_{\mu\alpha}={i\over2}\sigma_{\mu\alpha\dot\alpha}\bar\sigma^{\nu\dot\alpha\beta}S_{\nu\beta}+(\sigma_{\mu}^{\ \nu})_{\alpha}^{\ \beta}\partial_{\nu}\gamma_{\beta}.
}

Now, let us impose closure of $\CO_{\mu}$ under the SUSY transformations. To that end, we note that closure requires that
\eqn\closureh{\eqalign{
\left(\eta^{\beta}\xi^{\alpha}-\xi^{\beta}\eta^{\alpha}\right)\delta_{\beta}\delta_{\alpha}\CO_{\mu}&=0,\cr\left(\xi^{\alpha}\bar\eta_{\dot\alpha}\delta^{\dot\alpha}\delta_{\alpha}-\bar\eta_{\dot\alpha}\xi^{\alpha}\delta_{\alpha}\delta^{\dot\alpha}\right)\CO_{\mu}&=2i\xi\sigma^{\nu}\bar\eta\partial_{\nu}\CO_{\mu}.
}}
The first equation implies\foot{In what follows, we define $\xi^{\alpha}\delta_{\alpha}\CO=i\left[\xi Q, \CO\right]$ and $\bar\xi_{\dot\alpha}\delta^{\dot\alpha}\CO=i\left[\bar\xi\bar Q, \CO\right]$. Our conventions follow Wess and Bagger.}
\eqn\consclosi{
i\partial^{\nu}\gamma_{\nu\mu}-{1\over2}\epsilon_{\nu\mu\rho\lambda}\partial^{\nu}\gamma^{\rho\lambda}+{3\over2}\partial_{\mu}\gamma=2i\partial^{\nu}\gamma_{\nu\mu}+{3\over2}\partial_{\mu}\gamma=0,
}
where we have defined
\eqn\defgamma{
\delta_{\alpha}\gamma_{\beta}\equiv i\epsilon_{\beta\alpha}\gamma-\left(\sigma^{\mu\nu}\right)_{\alpha\beta}\gamma_{\mu\nu}.
}
In the second equality of \consclosi, we have used the self-duality of $\gamma_{\mu\nu}$. Taking the divergence of \consclosi\ and using the antisymmetry of $\gamma_{\mu\nu}$, we find that
\eqn\consclosii{
\partial^2\gamma=0.
}
Therefore, we see that $\gamma$ is a free scalar of dimension three. However, since our theory is scale-invariant, the unitarity bound \Scalarcond\ implies that $\gamma$ must have dimension one, and so
\eqn\consunit{
\gamma=0.
}
Applying this result to \consclosi\ we see that
\eqn\consclosii{
\partial^{\nu}\gamma_{\nu\mu}=0 \ \ \ \Rightarrow  \ \ \ \gamma_{\mu\nu}=0,
}
where the vanishing of $\gamma_{\mu\nu}$ follows from \Maxeqn\ and the fact that it has dimension three. Finally, from these results, it follows that $\gamma_{\alpha}$ is antichiral
\eqn\antichircond{
\left\{Q_{\beta}, \gamma_{\alpha}\right\}=0,
}
and
\eqn\consunitii{
D^2\CO_{\mu}=\bar D^2\CO_{\mu}=0.
}

Next, consider the second equation in \closureh. From this equation it follows that
\eqn\consclos{
\partial_{\nu}\CO_{\mu}={1\over4}\eta_{\mu\nu}D-{1\over8}\epsilon_{\rho\lambda\mu\nu}F^{\rho\lambda}-{1\over4}\partial_{\nu}\left(\gamma_{\mu}+\bar\gamma_{\mu}\right)+{1\over4}\eta_{\mu\nu}\partial^{\rho}\left(\gamma_{\rho}+\bar\gamma_{\rho}\right)-{i\over4}\epsilon_{\gamma\rho\mu\nu}\partial^{\gamma}\left(\gamma^{\rho}-\bar\gamma^{\rho}\right),
}
where we have defined
\eqn\gammamu{
\delta_{\dot\alpha}\gamma_{\alpha}\equiv\sigma^{\mu}_{\alpha\dot\alpha}\gamma_{\mu}.
}
We can solve for $\CO_{\mu}$ in terms of $\gamma_{\mu}$ as follows. First, consider taking the trace of \consclos
\eqn\trace{
\partial^{\mu}\CO_{\mu}=D+{3\over4}\partial^{\mu}\left(\gamma_{\mu}+\bar\gamma_{\mu}\right).
}
Recalling from \Dcond\ that $D=-4T=4\partial^{\mu}\CO_{\mu}$, we see that
\eqn\tracefinal{
\partial^{\mu}\CO_{\mu}=-{1\over4}\partial^{\mu}\left(\gamma_{\mu}+\bar\gamma_{\mu}\right).
}
Next, consider taking the divergence of \consclos\ with respect to $x^{\nu}$. We find
\eqn\divcond{
\partial^2\CO_{\mu}=-{1\over4}\partial^2\left(\gamma_{\mu}+\bar\gamma_{\mu}\right).
}
From \tracefinal\ it follows that
\eqn\constr{
\CO_{\mu}=-{1\over4}\left(\gamma_{\mu}+\bar\gamma_{\mu}\right)+\hat j_{\mu}, \ \ \ \partial^{\mu}\hat j_{\mu}=0.
}
Imposing \divcond, we then see that
\eqn\consjhat{
\partial^2\hat j_{\mu}=0.
}
This equation implies that the two-point function
\eqn\twopt{
\left<\hat j_{\mu}(x)\hat j_{\nu}(0)\right>=0,
}
and unitarity requires that $\hat j_{\mu}=0$. Therefore, we arrive at the conclusion that
\eqn\concCOmu{
\CO_{\mu}=-{1\over4}\left(\gamma_{\mu}+\bar\gamma_{\mu}\right).
}
Note that in superfield language, this equation can be written as
\eqn\concCOmuSF{
\CO_{\mu}={1\over8}\bar\sigma_{\mu}^{\dot\alpha\alpha}\left(\bar D_{\dot\alpha}\Gamma_{\alpha}-D_{\alpha}\bar\Gamma_{\dot\alpha}\right),
}
where $\gamma_{\alpha}\equiv\Gamma_{\alpha}|$. Also, using \concCOmu\ it follows from the vanishing of the antisymmetric part of \consclos\ that
\eqn\concCOmui{
U_{\mu}=-i\left(\gamma_{\mu}-\bar\gamma_{\mu}\right)+\partial_{\mu}\hat\CO,
}
where $U_{\mu}$ is the vector component of $U$ in \FZwelldef, and $\hat\CO$ is a local gauge-invariant scalar defined up to shifts by a constant.

To proceed further, we impose closure on $\gamma_{\alpha}$. In particular, consider imposing
\eqn\closgamma{
\left(\xi^{\alpha}\bar\eta_{\dot\alpha}\delta^{\dot\alpha}\delta_{\alpha}-\bar\eta_{\dot\alpha}\xi^{\alpha}\delta_{\alpha}\delta^{\dot\alpha}\right)\gamma_{\beta}=2i\xi^{\alpha}\bar\eta^{\dot\alpha}\sigma^{\nu}_{\alpha\dot\alpha}\partial_{\nu}\gamma_{\beta}.
}
A bit of algebra reveals that
\eqn\Diraceqn{
\sigma^{\mu}_{\alpha\dot\alpha}\partial_{\mu}\left(\gamma^{\alpha}+{1\over2}U_{\alpha}-{i\over2}\hat\CO_{\alpha}\right)=0,
}
where $\hat\CO_{\alpha}\equiv\left[Q_{\alpha}, \hat\CO\right]$ is a well-defined operator.\foot{In particular, the ambiguity in the definition of $\hat\CO$ corresponding to shifts by a constant is annihilated by the supercharge.} Since the fermions in \Diraceqn\ all have dimension $5/2$, the unitarity bound in \Diraccond\ forces
\eqn\unitDirac{
\gamma_{\alpha}=-{1\over2}U_{\alpha}+{i\over2}\hat\CO_{\alpha}.
}
Translating to superfield language this implies that $\Gamma_{\alpha}={i\over2}D_{\alpha}U+{1\over2}D_{\alpha}\hat\CO$ and so
\eqn\COfinal{
\CO_{\mu}=-{1\over4}\partial_{\mu}U+{1\over16}\bar\sigma_{\mu}^{\dot\alpha\alpha}\left[\bar D_{\dot\alpha}, D_{\alpha}\right]\hat\CO.
}

\subsec{From scale invariance to superconformality}
In the previous subsection we discovered that unitarity and scale invariance imposed strong constraints on the virial current multiplet, $\CO_{\mu}$, of an R-symmetric fixed point. In particular, we discovered that
\eqn\COsummary{\eqalign{
&D^2\CO_{\mu}=\bar D^2\CO_{\mu}=0,\cr&\CO_{\mu}=-{1\over4}\partial_{\mu}U+{1\over16}\bar\sigma_{\mu}^{\dot\alpha\alpha}\left[\bar D_{\dot\alpha}, D_{\alpha}\right]\hat\CO,
}}
where $\hat\CO$ is defined up to possible shifts by a constant. In this section, we will use these results to describe a class of unitary R-symmetric fixed points that are necessarily superconformal.

To proceed, notice that the second equation in \COsummary\ implies that $\CO_{\mu}$ cannot mix with vector primaries\foot{We should point out that by a primary we mean an operator that cannot be written as the supercharge variation of a well-defined lower dimensional operator.}, or, more generally, with vector operators, $\tilde\CO_{\mu}$, that do not have well-defined SUSY partners of the form $\partial_{\mu}\tilde\CO$ . Therefore, we conclude that a theory cannot have only one nonconserved dimension two SSO (modulo deformations by conserved SSOs).  Indeed, either $U$ is such a nonconserved SSO, or the theory is superconformal. However, if $U$ is nonconserved, then the first equation in \COsummary\ implies that
\eqn\SCres{
D^2U=0,
}
and so we arrive at a contradiction and see that the theory is necessarily a SCFT. In fact, we can arrange for the superconformal anomaly to vanish by making a transformation of the form \Ramb, \viramb.

The status of theories with more than one nonconserved dimension two SSO is unclear since in that case \COsummary\ is only enough to conclude that $U$ is part of a left-conserved (annihilated by $\bar D^2$) superfield or a right-conserved (annihilated by $D^2$) superfield. We hope to analyze such theories in the near future. However, we will see below that if the fixed point under consideration arises as the end point of an RG flow from a UV SCFT satisfying certain simple properties, then we will be able to conclude that the IR fixed point is necessarily conformal (in spite of not knowing the precise operator spectrum).

\subsec{A first look at the IR behavior of SQCD}
The above discussion is enough to show that there cannot be any nonconformal perturbative fixed points in SQCD (and a large class of related theories). Indeed, consider $SU(N_c)$ $\CN=1$ SQCD in the conformal window, i.e. with ${3\over2}N_c<N_f<3N_c$. The charged matter spectrum then consists of chiral squark superfields $Q$ and $\tilde Q$ transforming as follows under the symmetries of the theory
\eqn\tableone{\matrix{& SU(N_c) & SU(N_f)\times SU(N_f) & U(1)_R & U(1)_B & U(1)_A\cr & \cr  Q & {\bf N_c} & {\bf N_f\times 1}& 1-{N_c\over N_f} & 1 & 1 \cr \tilde Q & {\bf \bar N_c} & {\bf 1\times N_f}& 1-{N_c\over N_f} & -1 & 1    }}
All of the symmetries in the above table are good quantum symmetries with the exception of the axial symmetry, $U(1)_A$, which suffers from an anomaly. 

Using the above degrees of freedom, we would like to construct dimension two SSO contributions to the virial superfield consistent with the general form we have described in \COsummary. In particular, the SSOs must be singlets under $SU(N_c)\times SU(N_f)^2\times U(1)_R\times U(1)_B$. Clearly there cannot be any contributions from the baryons and mesons of the theory. In fact, the only objects that can contribute are the $U(1)_B$ and $U(1)_A$ current superfields. In particular, since we assume that $U$ is nonconserved, \COsummary\ must take the form
\eqn\COIRSQCDpert{
\CO_{\mu}=-{1\over4}\partial_{\mu}U+a_1\bar\sigma_{\mu}^{\dot\alpha\alpha}\left[\bar D_{\dot\alpha}, D_{\alpha}\right]U+a_2\bar\sigma_{\mu}^{\dot\alpha\alpha}\left[\bar D_{\dot\alpha}, D_{\alpha}\right]J_{B}
}
where $J_B$ is the conserved $U(1)_B$ current superfield. From the discussion around \SCres\ we see that
\eqn\SCQCDpert{
D^2U=0.
}
Therefore, $U$ must be conserved, and the putative IR fixed point is superconformal as promised.

This discussion generalizes Polchinski's results regarding the conformality of non-SUSY BZ fixed points consisting of gauge fields coupled to fermions.

\subsec{The RG flow and constraints from the UV in SQCD-like theories}
As we just saw, all unitary R-symmetric fixed points are either superconformal or have more than one nonconserved dimension two SSO (modulo deformations by a conserved SSO). This fact allowed us to conclude that all perturbative SQCD fixed points are necessarily superconformal.

In this section we will argue that SUSY still allows us to show conformality of the IR fixed points of SQCD-like theories even when we do not have detailed knowledge of the IR spectrum of operators. To see this, we will consider an RG flow from a weakly-coupled UV SCFT fixed point to a potentially strongly-coupled fixed point in the IR. We will see that as long as the flow is initiated by a marginally relevant perturbation that leads to only one dimension two nonconserved SSO in the UV, and as long as the UV theory has a nonanomalous R-symmetry and a well-defined FZ multiplet, the IR fixed point is necessarily superconformal.

The crucial point is that the decomposition of the virial current superfield in \COsummary\ is well-defined. In particular, there cannot be any cancellation between the first term, proportional to $\partial_{\mu}U$, and the second term, proportional to $\left[\bar D_{\dot\alpha}, D_{\alpha}\right]\hat\CO$ (i.e., the two terms transform in different representations of SUSY). Indeed, any such cancellation would imply that $\chi_{\alpha}\sim \bar D^2D_{\alpha}U=0$, and the theory would then be superconformal.

Furthermore, we can always choose the operators appearing in \COsummary\ so that $U$ descends from a UV operator of dimension two. As a result, we see that under a marginally relevant deformation in the UV, the virial current superfield can be written as follows (up to operators that vanish in the IR)
\eqn\virUV{
\CO_{\mu}^{UV}=-{1\over4}\Lambda^{D-3}\partial_{\mu}U^{UV}+c_2\Lambda^{D-d}\bar\sigma_{\mu}^{\dot\alpha\alpha}\left[\bar D_{\dot\alpha}, D_{\alpha}\right]\hat\CO^{UV},
}
where $D$ is the UV dimension of $\CO_{\mu}$, and $d-1\ge2$ is the dimension of $\hat\CO^{UV}$. 

Perturbativity in the UV forces $D=d=3$ (this conclusion also follows from the fact that $\CO_{\mu}$ cannot be the $\theta\bar\theta$ component of the $\hat\CO$ superfield only up to some finite order in $\Lambda$), and so we can write an operator equation for $\CO_{\mu}$ in the UV of the following form
\eqn\COUV{
\CO_{\mu}^{UV}=-{1\over4}\partial_{\mu}U^{UV}+a_1\bar\sigma_{\mu}^{\dot\alpha\alpha}\left[\bar D_{\dot\alpha}, D_{\alpha}\right]U^{UV}+\sum_{i=2,N}a_i\bar\sigma_{\mu}^{\dot\alpha\alpha}\left[\bar D_{\dot\alpha}, D_{\alpha}\right]J^{i, \ UV}
}
where the $J^i$ are the dimension two conserved SSOs (we assume there is only one nonconserved SSO of dimension two, which must correspond to $U^{UV}$).

Now, flowing to the IR, we encounter an operator of the form \COsummary, the discussion around \SCres\ applies, and in the IR we have
\eqn\COIRconseq{
D^2U=0.
}
Therefore, the IR fixed point is superconformal. In particular, note that this discussion applies to SQCD in the conformal window (i.e., ${3\over2}N_c< N_f<3N_c$), and so we conclude that the conformal window is actually conformal.

\newsec{Conclusions and open problems}
We have shown that a scale-invariant, unitary, R-symmetric theory is either superconformal or it has at least two nonconserved dimension two SSOs. Furthermore, we generalized this result and showed that the IR fixed points of UV SCFTs deformed by marginally relevant perturbations that leave only one dimension two nonconserved SSO in the UV are also superconformal provided the RG flow is R-symmetric and the UV theory has an FZ multiplet. This allowed us to conclude that conformal window of SQCD is indeed conformal.

There are several interesting potential extensions of our work. First, we would like to try to understand what happens for a more general R-symmetric fixed point where we allow for potentially many nonconserved dimension two SSOs without assuming that the theory is the end point of an RG flow of the type described above. It may well be that such theories are necessarily conformal. A sufficient result to prove this statement would be to show that in any unitary R-symmetric fixed point, left conserved superfields (those annihilated by $\bar D^2$) are necessarily right conserved as well.

Second, we assumed the existence of an R-symmetry. While such symmetries are typically present in examples of physical interest (like SQCD), we could imagine generalizing our methods to cases where R-symmetry is not assumed.

At a more abstract level, it would also be interesting to understand any potential connection between our results and the existence of a four-dimensional version of the c-theorem (perhaps, along the lines of the conjectured \lq\lq a-theorem" \CardyCWA). Indeed, as we remarked in the introduction, Zamolodchikov's and Polchinski's proof of the fact that scale invariance implied full conformal invariance used the c-theorem. Of course, it is possible to deduce this fact independently of the existence of a c-function, and a corresponding four-dimensional quantity did not appear in our proof, but there seems to be some connection between the two concepts.\foot{This connection has been further hinted at on the gravity side by Nakayama \NakayamaWX.} We hope to return to these questions in the near future.

\bigskip
\bigskip\centerline{\bf Acknowledgements}
We are grateful to S. Ferrara, N. Seiberg, and, especially, Z. Komargodski for interesting comments and discussions. This work was supported in part by the European Commission under the ERC Advanced Grant 226371 and the contract PITN-GA-2009-237920. I. A. was also supported in part by the CNRS grant GRC APIC PICS 3747.

\vfill\eject
\appendix{A}{Unitarity constraints}
In this appendix we will derive the unitarity constraints in \fermunit, which we reproduce below for ease of reference
\eqn\funitapp{
\partial^{\nu}\sigma_{\nu\mu}^{\alpha\beta}\CO_{\alpha\beta\delta}=\partial^{\nu}\bar\sigma_{\nu\mu}^{\dot\alpha\dot\beta}\CO_{\dot\alpha\dot\beta\delta}=0,
}
where $\CO_{\alpha\beta\gamma}$ and $\CO_{\dot\alpha\dot\beta\delta}$ are dimension $5/2$ operators transforming with spin $(3/2,0)$ and $(1/2,1)$ respectively.

Let us first consider the totally symmetric operator $\CO_{\alpha\beta\delta}$. Using Lorentz invariance and spin index symmetry, we see that the two-point function of this operator with its conjugate must have the following form
\eqn\COthtw{
\left<\CO_{\alpha\beta\delta}(x)\bar\CO_{\dot\gamma\dot\epsilon\dot\eta}(0)\right>={1\over(2\pi)^2}f_{\alpha\beta\delta\dot\gamma\dot\epsilon\dot\eta}^{\rho_1\rho_2\rho_3}\partial_{\rho_1}\partial_{\rho_2}\partial_{\rho_3}\left({1\over x^2}\right),
}
where the tensor $f$ is built from dimensionless constants. Now, consider contracting $\CO_{\alpha\beta\delta}$ ($\bar\CO_{\dot\gamma\dot\epsilon\dot\eta}$) with $\partial^{\nu_1}\sigma_{\nu_1\mu_1}^{\alpha\beta}$ ($\partial^{\nu_2}\bar\sigma_{\nu_2\mu_2}^{\dot\gamma\dot\epsilon}$) and computing the two-point function
\eqn\twopttwo{
\left<\partial^{\nu_1}\sigma_{\nu_1\mu_1}^{\alpha\beta}\CO_{\alpha\beta\delta}(x)\partial^{\nu_2}\bar\sigma_{\nu_2\mu_2}^{\dot\gamma\dot\epsilon}\bar\CO_{\dot\gamma\dot\epsilon\dot\eta}(0)\right>.
}
This two-point function involves five derivatives and has only three free vector indices. Therefore, in order to have a non-trivial result, at least one pair of derivatives must be contracted against each other yielding a Laplacian acting on $x^{-2}$. As a result, we see that
\eqn\COthtwi{
\left<\partial^{\nu_1}\sigma_{\nu_1\mu_1}^{\alpha\beta}\CO_{\alpha\beta\delta}(x)\partial^{\nu_2}\bar\sigma_{\nu_2\mu_2}^{\dot\gamma\dot\epsilon}\bar\CO_{\dot\gamma\dot\epsilon\dot\eta}(0)\right>=0, \ \ \ x\ne0.
}
Unitarity then implies that
\eqn\unitconsi{
\partial^{\nu}\sigma_{\nu\mu}^{\alpha\beta}\CO_{\alpha\beta\delta}=0.
}

Next, consider the operator $\CO_{\dot\alpha\dot\beta\delta}$. Lorentz invariance and spin index symmetry imply that the two point function of this operator with its conjugate satisfies
\eqn\COh{\eqalign{
\langle\CO_{\dot\alpha\dot\beta\delta}(x)\bar\CO_{\gamma\epsilon\dot\eta}(0)\rangle&={a_1\over (2\pi)^2}\left(\epsilon_{\delta\gamma}\epsilon_{\dot\alpha\dot\eta}\sigma^{\mu}_{\epsilon\dot\beta}+\epsilon_{\delta\epsilon}\epsilon_{\dot\alpha\dot\eta}\sigma^{\mu}_{\gamma\dot\beta}+\epsilon_{\delta\gamma}\epsilon_{\dot\beta\dot\eta}\sigma^{\mu}_{\epsilon\dot\alpha}+\epsilon_{\delta\epsilon}\epsilon_{\dot\beta\dot\eta}\sigma^{\mu}_{\gamma\dot\alpha}\right)\partial_{\mu}\left({1\over x^4}\right)\cr&+{1\over(2\pi)^2}\tilde f_{\dot\alpha\dot\beta\delta\gamma\epsilon\dot\eta}^{\rho_1\rho_2\rho_3}\partial_{\rho_1}\partial_{\rho_2}\partial_{\rho_3}\left({1\over x^2}\right)}.
}
By the same reasoning as above, the term with three derivatives will yield only contact terms when the operators in the correlation function are acted upon by derivatives. Furthermore, as we will now show, unitarity forces $a_1=0$.

To derive the constraints imposed by unitarity, we follow \GrinsteinQK\ and consider scattering of particles, $\chi^{\dot\alpha\dot\beta\delta}$, coupled to the fixed-point operators via
\eqn\coupling{
\CL\supset\chi^{\dot\alpha\dot\beta\delta}\CO_{\dot\alpha\dot\beta\delta}+{\rm h.c.}
}
Using the formula
\eqn\ft{
{1\over(2\pi)^2}{1\over (x^2)^d}={\Gamma(2-d)\over4^{d-1}\Gamma(d)}\int{d^4k\over(2\pi)^4}e^{ikx}(k^2)^{d-2},
}
we find the following amplitude for $\chi\to\bar\chi$ scattering
\eqn\amp{\eqalign{
\CA=&a_1\chi^{\dot\alpha\dot\beta\delta}\left(\epsilon_{\delta\gamma}\epsilon_{\dot\alpha\dot\eta}\sigma^{\mu}_{\epsilon\dot\beta}+\epsilon_{\delta\epsilon}\epsilon_{\dot\alpha\dot\eta}\sigma^{\mu}_{\gamma\dot\beta}+\epsilon_{\delta\gamma}\epsilon_{\dot\beta\dot\eta}\sigma^{\mu}_{\epsilon\dot\alpha}+\epsilon_{\delta\epsilon}\epsilon_{\dot\beta\dot\eta}\sigma^{\mu}_{\gamma\dot\alpha}\right)\bar\chi^{\gamma\epsilon\dot\eta}k_{\mu}{\Gamma(5/2-d)\over 4^{d-3/2}\Gamma(d-1/2)}\cr&\cdot(-k^2-i\epsilon)^{d-5/2}-\chi^{\dot\alpha\dot\beta\delta}\tilde f^{\rho_1\rho_2\rho_3}_{\dot\alpha\dot\beta\delta\gamma\epsilon\dot\eta}\bar\chi^{\gamma\epsilon\dot\eta}k_{\rho_1}k_{\rho_2}k_{\rho_3}{\Gamma(7/2-d)\over4^{d-5/2}\Gamma(d-3/2)}\cdot(-k^2-i\epsilon)^{d-7/2},
}}
where $d=5/2$ is the dimension of $\CO_{\dot\alpha\dot\beta\delta}$.

Unitarity, in the guise of the optical theorem, requires that the imaginary part of the forward scattering amplitude be positive semidefinite. Using the formula
\eqn\helpform{
\Gamma(1-x)\Gamma(x)\sin(\pi x)=\pi,
}
we find
\eqn\ampfwd{\eqalign{
{\rm Im} \ \CA_{\rm fwd}&=(k^2)^{d-5/2}{\pi(d-3/2)\over 4^{d-3/2}\Gamma^2(d-1/2)}\Big[a_1\chi^{\dot\alpha\dot\beta\delta}\Big(\epsilon_{\delta\gamma}\epsilon_{\dot\alpha\dot\eta}\sigma^{\mu}_{\epsilon\dot\beta}+\epsilon_{\delta\epsilon}\epsilon_{\dot\alpha\dot\eta}\sigma^{\mu}_{\gamma\dot\beta}+\epsilon_{\delta\gamma}\epsilon_{\dot\beta\dot\eta}\sigma^{\mu}_{\epsilon\dot\alpha}\cr&+\epsilon_{\delta\epsilon}\epsilon_{\dot\beta\dot\eta}\sigma^{\mu}_{\gamma\dot\alpha}\Big)\bar\chi^{\gamma\epsilon\dot\eta}k_{\mu}+4(5/2-d)(d-3/2)\chi^{\dot\alpha\dot\beta\delta}\tilde f^{\rho_1\rho_2\rho_3}_{\dot\alpha\dot\beta\delta\gamma\epsilon\dot\eta}\bar\chi^{\gamma\epsilon\dot\eta}k^{-2}\Big]\theta(k^0)\theta(k^2).
}}
For $d=5/2$, we see that the term proportional to $\tilde f$ vanishes leaving only the term proportional to $a_1$. Going to the rest frame $k_{\mu}=k_0>0$ and arranging a wave packet with only $\chi^{212}=\chi^{122}\ne0$ and $\chi^{111}\ne0$, we see that
\eqn\ampfwdwvi{
{\rm Im}\ \CA_{\rm fwd}={\pi k_0a_1\over2}\left(2|\chi^{212}|^2-\chi^{111}\bar\chi^{212}-\chi^{212}\bar\chi^{111}\right).
}
It is now easy to see that if $a_1\ne0$, \ampfwdwvi\ has an indefinite sign. Indeed, suppose we take $\chi^{212}$ real and positive. Setting $\chi^{111}=0$ and taking $a_1>0$ without loss of generality, we find that \ampfwdwvi\ has positive sign. On the other hand, taking $\chi^{111}$ real and sufficiently positive, we see that \ampfwdwvi\ has negative sign. Therefore, we conclude that $a_1=0$ and so by the same reasoning that led to \unitconsi, we find
\eqn\finalconclusion{
\partial^{\nu}\bar\sigma_{\nu\mu}^{\dot\alpha\dot\beta}\CO_{\dot\alpha\dot\beta\delta}=0.
}

\bigskip\bigskip
\appendix{B}{Scaling of the supercurrent multiplet}
Before concluding, we should note that throughout our paper, as well as in the proof of the unitarity constraint \fermunit\ in the appendix, we have assumed that the supercurrent multiplet has canonical scaling. In general, however, the commutator of the R-current with the scale charge has a Schwinger term, $\partial^{\nu}A_{\nu\mu}$. Such a term leads to noncanonical scaling of the whole supercurrent multiplet since
\eqn\noncanon{
\left[ j_{\mu}^R, \Delta\right]=-ix^{\rho}\partial_{\rho}j_{\mu}^R-3ij_{\mu}^R+i\partial^{\nu}A_{\nu\mu},
}
where
\eqn\Adef{
A_{\nu\mu}=-A_{\mu\nu},
}
without affecting the resulting R-charge dimension itself
\eqn\Rchargecomm{
\left[R, \Delta\right]=0.
}
However, we are always free to define a new, conserved, R-current, $\hat j_{\mu}^R$, that satisfies a canonical scaling relation
\eqn\canon{
\left[ \hat j_{\mu}^R, \Delta\right]=-ix^{\rho}\partial_{\rho}\hat j_{\mu}^R-3i\hat j_{\mu}^R.
}

To see that we can always define a current such that \canon\ indeed holds, we will generalize Polchinski's argument in \PolchinskiDY\ to the present case. To that end, consider a complete set of antisymmetric operators $A^a_{\mu\nu}$ such that
\eqn\completeness{
A_{\nu\mu}=y_aA^a_{\nu\mu}.
}
The operators on the RHS of the above equation satisfy the following scaling relations
\eqn\scalrelatAcomp{
\left[A_{\nu\mu}^a, \Delta\right]=-ix^{\rho}\partial_{\rho}A^a_{\nu\mu}-i(2+\omega)^a_{\ b}A^b_{\nu\mu},
}
where $\omega$ is a matrix with eigenvalues greater than or equal to zero. Now, consider defining
\eqn\impRcurr{
\hat j_{\mu}^R=j_{\mu}^R+y_{a}(\hat\omega^{-1})^{a}_{\ \  b}\partial^{\nu}A^b_{\nu\mu}.
}
Here $\hat\omega$ is the restriction of $\omega$ to a matrix of positive definite eigenvalues. This restriction is justified since antisymmetric operators of dimension two (i.e., with $\omega=0$ in \scalrelatAcomp) satisfy a Maxwell equation. Given this construction, we see that $\hat j_{\mu}^R$  is conserved, and that it also satisfies \canon. Note that the matrix inverse in \impRcurr\ is defined since the spectrum of $\hat\omega$ is positive. Furthermore, unlike in two dimensions, we do not expect any infrared subtleties in the two-point functions of the current multiplet.

\listrefs
\end